 \documentclass{easychair}

\usepackage{doc}
\usepackage{float}
\usepackage{amsfonts}
\usepackage{caption}
\usepackage{subcaption}

\title{Emergent Crossing Regimes of Identical Autonomous Vehicles at an Uncontrolled Intersection}

\author{
Karam Safarov\inst{1}
\and
    Thomas Kent\inst{1}
\and
   Eddie Wilson\inst{1}
\and
   Arthur Richards\inst{1}
}

\institute{Faculty of Engineering,
  University of Bristol,
  Bristol, UK\\
  \email{karam.safarov@bristol.ac.uk}}

\authorrunning{Safarov, Kent, Wilson and Richards}

\titlerunning{Crossing Regimes at an Intersection}

\begin{document}

\maketitle
\section{Introduction}
\label{sect:introduction}
To investigate the impact of Autonomous Vehicles (AVs) on urban congestion, this study looks at their performance at road intersections.  Some suggest that because of increased environmental awareness, sensing and communication, AVs have the potential to outperform their human counterparts~\cite{Zohdy2016}. Others suggest that AVs might behave in simplistic ways and, sticking rigidly to the rules, make congestion worse~\cite{LeVine2015}.  Intersection performance will be a key determining factor: although human-drivers are typically very good at making decisions \cite{VanDenBeukel2011}, over one third of road accidents happen at the intersections \cite{Fortelle2015}.

Intersection performance has been studied across a range of traffic densities using a simple MATLAB simulation of two intersecting 1-D flows of homogeneous automated vehicles. This lacks the detail of more advanced simulations, such as those using VISSIM (Verkehr In Städten - SIMulationsmodell)~\cite{Atkins2016a,Li2013} or SUMO (Simulation of Urban MObility)~\cite{Krajzewicz2012,Qian2015}, but it enables fast identification of fundamental behaviours.  The results show that there are \emph{distinct crossing regimes} at low, medium and high densities. Furthermore, the transitions between regimes can be predicted analytically and their performance related to the fundamental model of 1-D traffic flow.  These findings have the potential to focus efforts on the development of improved decision-making rules for emerging AVs.
\section{Method}
\label{sect:methods}
\begin{figure}[t]
\centering
\begin{subfigure}{.43\textwidth}
 \includegraphics[width=1\columnwidth]{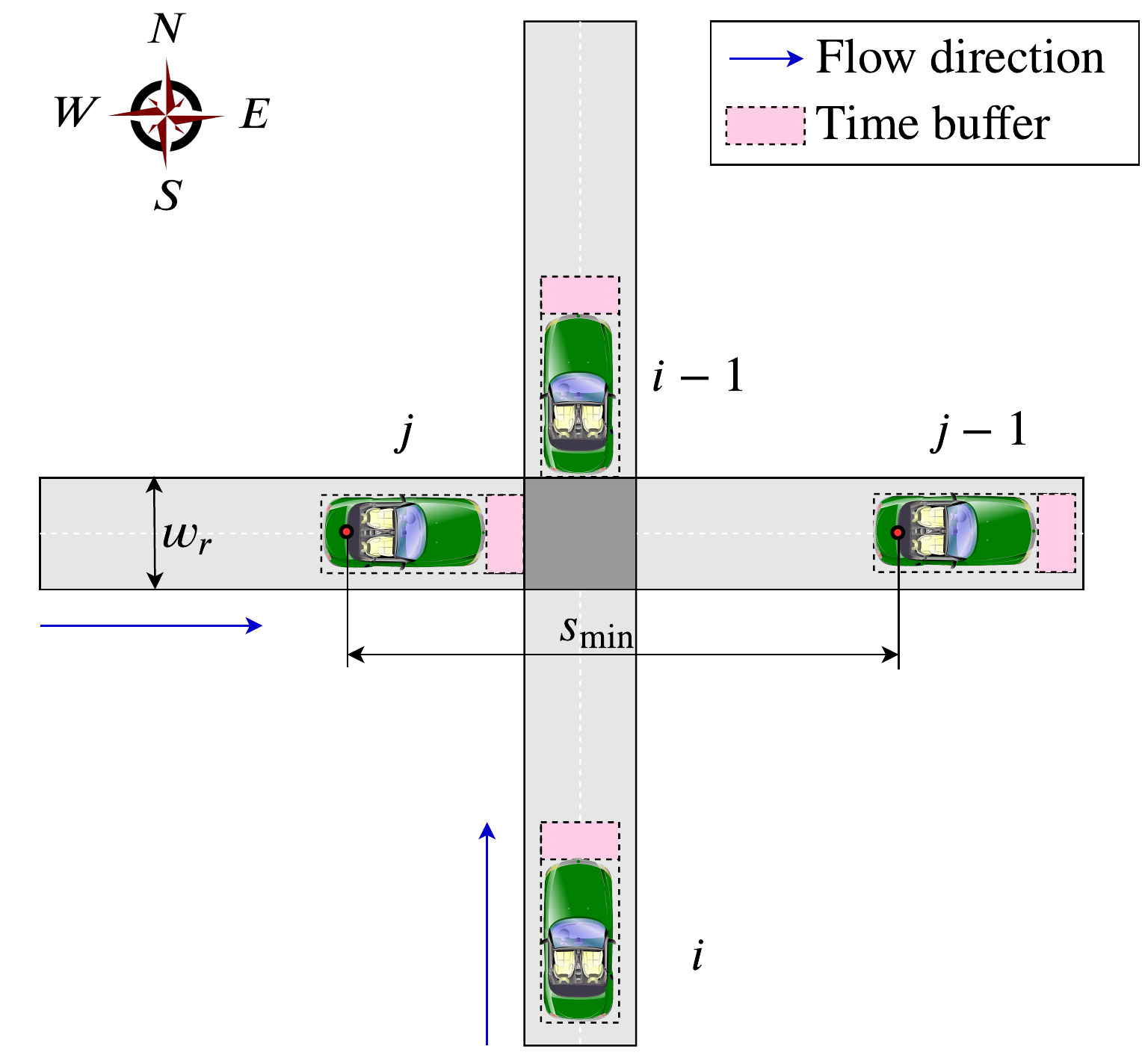}
\subcaption{Junction crossing at limiting point}
\label{fig:symmetric-turn-taking}
\end{subfigure}
\begin{subfigure}{.54\textwidth}
\includegraphics[width=1\columnwidth]{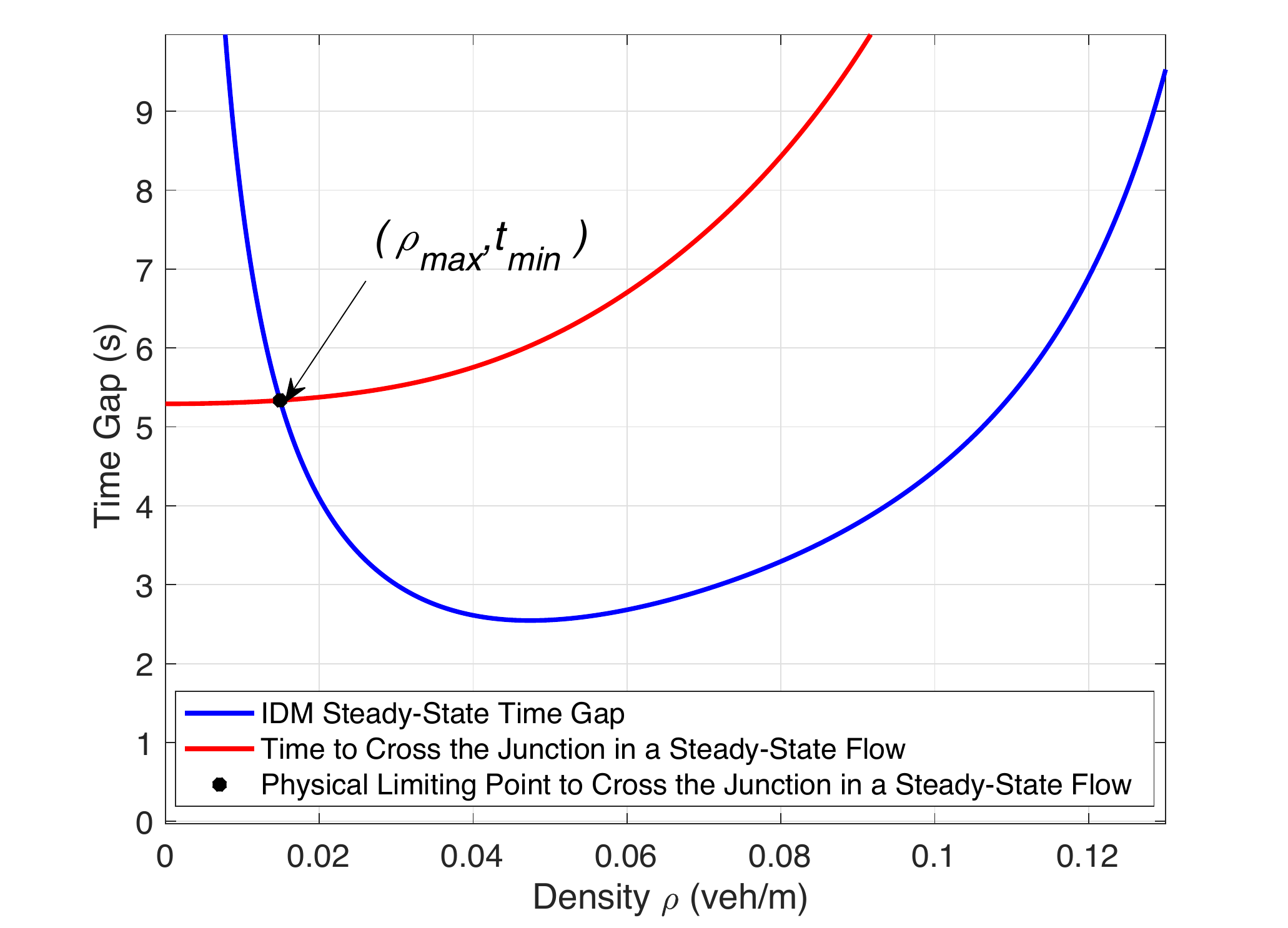}
\subcaption{Analytical limiting point $(\rho_{\text{max}}, t_{\text{min}})$}
\label{fig:analytic_critical}
\end{subfigure}
\caption{Physical limit of phased steady-state junction crossing schematics and analytical curves}
\end{figure}

We simulate two virtual ring-roads, each with a fixed number of vehicles~$n_c=30$, and vary the length~$L$ to enable a sweep of density~$\rho=n_c/L \in [0.002,0.130]\,\mathrm{veh/m}$ in steps of $0.001\,\mathrm{veh/m}$.  The two roads are identical and cross at a single intersection in the middle (Figure \ref{fig:symmetric-turn-taking}). Each simulation has 36000 time steps of size $\Delta t = 0.1\,\mathrm{s}$, which corresponds to one hour of real time. 

Away from the intersection, the vehicles use the Intelligent Driver Model (IDM) for collision-free car-following model using standard parameters~\cite{Treiber2013} of vehicle length $l = 4.4\,\mathrm{m}$, desired speed $v_0 = 13\,\mathrm{m/s}$, time gap $T = 1.6\,\mathrm{s}$ and acceleration $a = 1.5\,\mathrm{m/s^2}$
for which uniform flows are string-stable throughout the whole density sweep~\cite{Wilson2011}.

At the junction, vehicles decide to wait or proceed using a simple gap acceptance rules, crossing only when there is sufficient time to do so.  Crossing time includes a two-second buffer in advance of each vehicle (see Figure~\ref{fig:symmetric-turn-taking}). This is done for two reasons: (1) to guarantee a collision free crossing and (2) passenger comfort. 

\begin{figure}[t!]
\centering
\includegraphics[width=1\columnwidth]{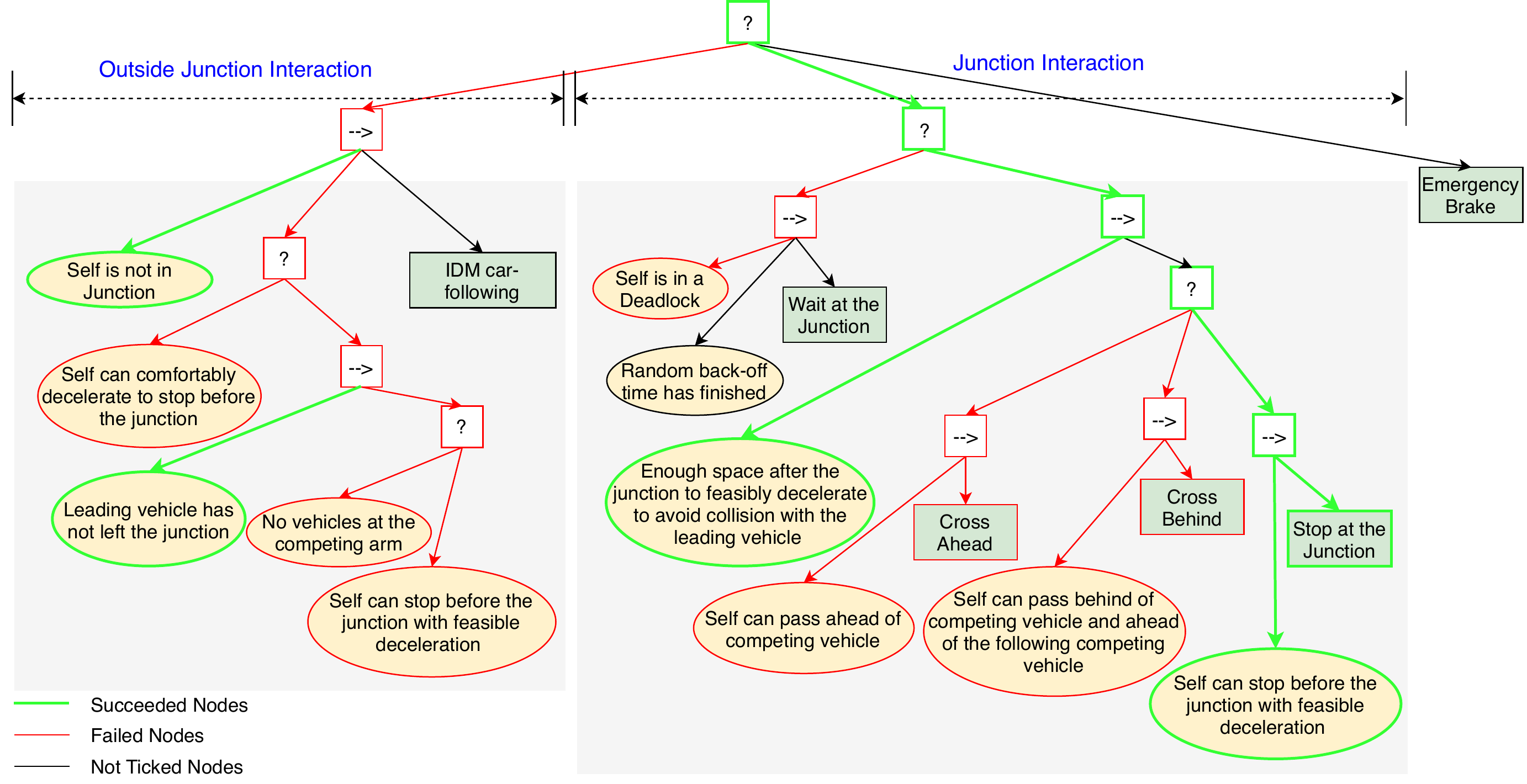}
\caption{Behaviour Tree based controller}
\label{fig:BT-a-tree}
\end{figure}

For modularity and future extensibility, each vehicle's  decision-making, including IDM and the crossing rules, is implemented in a Behaviour Tree (BT)~\cite{Colledanchise2017} (see Figure~\ref{fig:BT-a-tree}). The results of a single BT update are highlighted in the figure.  The first priority is the ``Outside Junction Interaction" behaviour, which fails in this case (red arrow), and flow moves to the ``Junction Interaction" behaviour.  Inside, the ``Cross Ahead" and the ``Cross Behind" behaviours both fail, so the lowest priority ``Stop at junction" action is triggered.  This succeeds (green arrow) so the whole ``Junction Interaction" module succeeds and there is no need to revert to the ``Emergency Brake" fall-back.

\begin{figure}[h]
\centering \includegraphics[width=0.6\columnwidth]{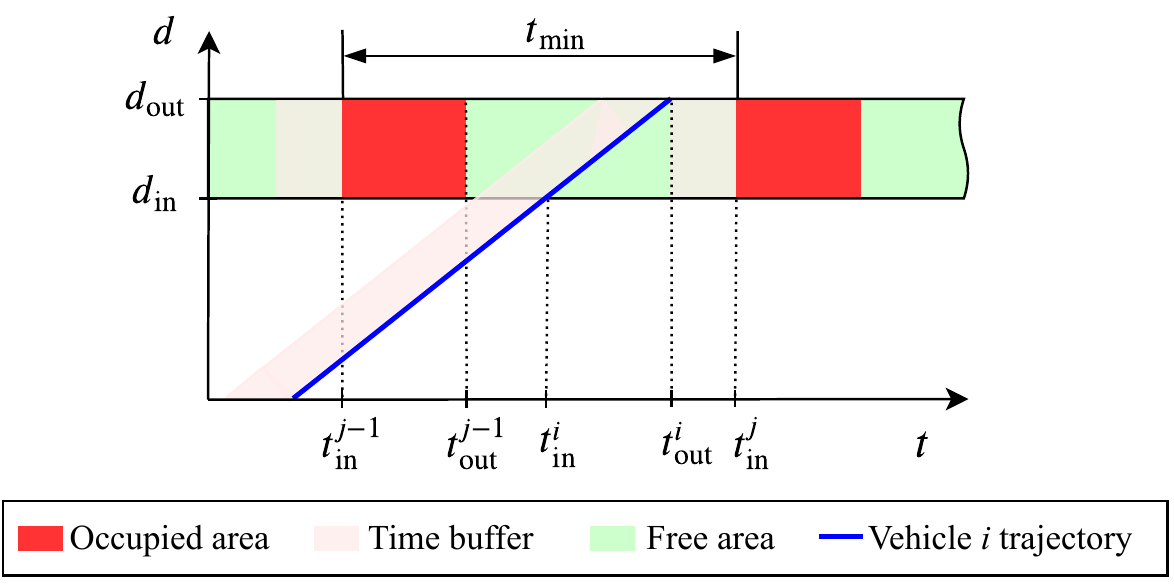}
\caption{Spatio-temporal trajectory with obstacles and buffers at the junction}
\label{fig:gap-accept}
\end{figure}

The simulations start with the vehicles positioned in equilibrium spacing and one road shifted by half that spacing relative to the other (phased crossing). This aims for the maximum throughput when the junction controller is not required for avoidance (as shown in the Figure~\ref{fig:symmetric-turn-taking}). The diagram in Figure~\ref{fig:gap-accept} shows an instance of this scenario, where the vehicle $i$ follows the vehicle $i-1$ and crosses the junction without collision. 
Since all traffic is homogeneous, junction arms are identical and density is fixed for the whole simulation, it is possible to calculate analytically $\rho_{\text{max}}$ (shown in Figure~\ref{fig:analytic_critical}) required for theoretical maximum throughput. A motivation of this paper is to demonstrate through simulation how these homogeneous flows of vehicles can achieve and sustain these maximum capacity regimes all the way to the density $\rho_{\text{max}}$. Moreover, $\rho_{\text{max}}$ corresponds to a spatial gap $s_{\text{min}}$ between vehicle $j$ and $j-1$ (shown in Figure~\ref{fig:symmetric-turn-taking}) that is a required \emph{space gap} for crossing the junction without the need for the junction logic to be \emph{activated}. Thus, for densities greater than $\rho_{\text{max}}$ the vehicles' space-gaps become too short for junction ``invisible” crossing and in turn the steady-state flow is no longer possible. 

Equivalently, the \emph{minimum time gap} can be formulated as $t_{\text{min}} = t_{\text{in}}^{j-1} - t_{\text{in}}^{j} = (t_{\text{out}}^{i} - t_{\text{in}}^i+ t_{\text{buff}}) +(t_{\text{out}}^{j-1} - t_{\text{in}}^{j-1} + t_{\text{buff}})$. Knowing that the steady-state parameters of both arms of the junction are identical, then $(t_{\text{out}}^{i} - t_{\text{in}}^i+ t_{\text{buff}}) = (t_{\text{out}}^{j-1} - t_{\text{in}}^{j-1} + t_{\text{buff}})$ and $v_e^{i} = v_e^{j-1} = v_e$, the equilibrium velocity. Next, we expand the equation $s_{\text{min}} = v_e t_{\text{min}}$ into $s_{\text{min}} = 2(d_{\text{out}} - d_{\text{in}}) + 2v_e t_{\text{buff}}$, where $(d_{\text{out}} - d_{\text{in}}) = l+w_r$ for the vehicle length $l$ and road width $w_r$ (Figure \ref{fig:symmetric-turn-taking}). The physical density limit can be calculated as $\rho_{\text{max}} = 1/{s_{\text{min}}}$, which for the vehicle length $4.4\,\mathrm{m}$ and road width $4\,\mathrm{m}$ of this paper results in $\rho_{\text{max}} = 0.01492\,\mathrm{veh/m}$. This value also corresponds to the point at which the \emph{IDM Steady-State time gap} intersects with the \emph{time to cross the junction in Steady-State flow}, as shown in Figure \ref{fig:analytic_critical}.
\section{Results}\label{sect:results}
\begin{figure}[h]
\centering
    \begin{subfigure}{1\textwidth}
        \centering
        \includegraphics[width=1\textwidth]{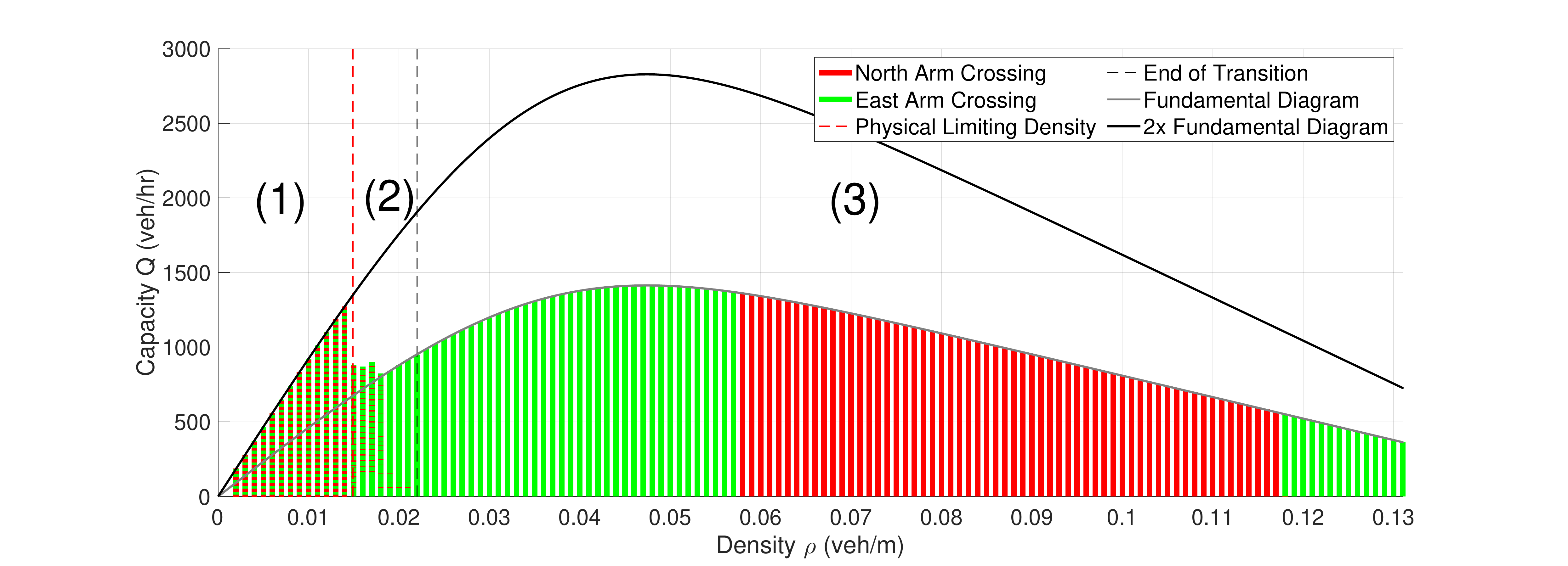}
        \vspace{-0.25in}
        \subcaption{Total throughput versus capacity: three emergent regimes} 
        \label{fig:a_fig}
    \end{subfigure}
    \begin{subfigure}{.32\textwidth}
        \includegraphics[width=1\textwidth]{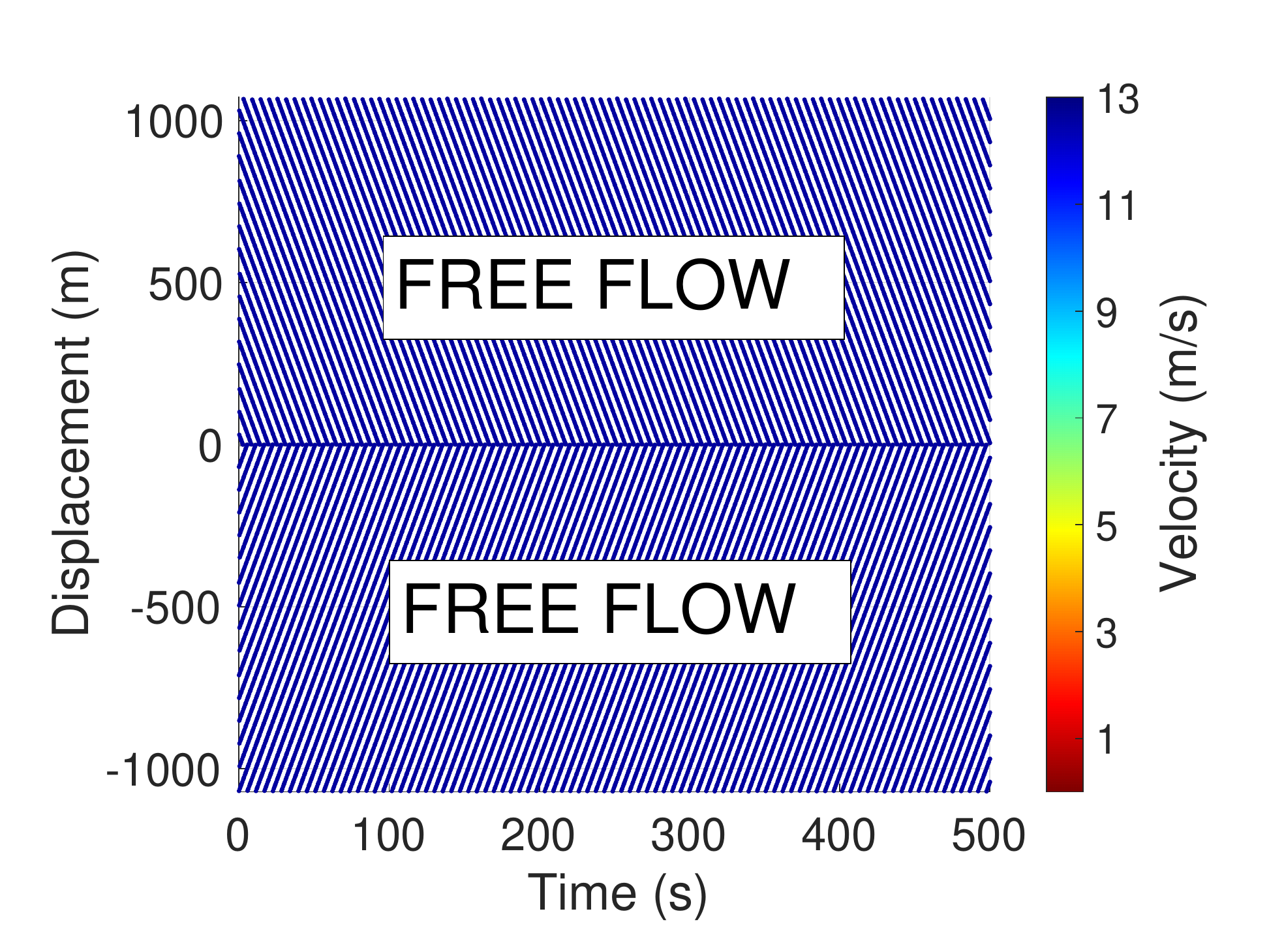}
        \subcaption{Regime (1)}
        \label{fig:b_fig}
    \end{subfigure}
    \begin{subfigure}{.32\textwidth}
        \includegraphics[width=1\textwidth]{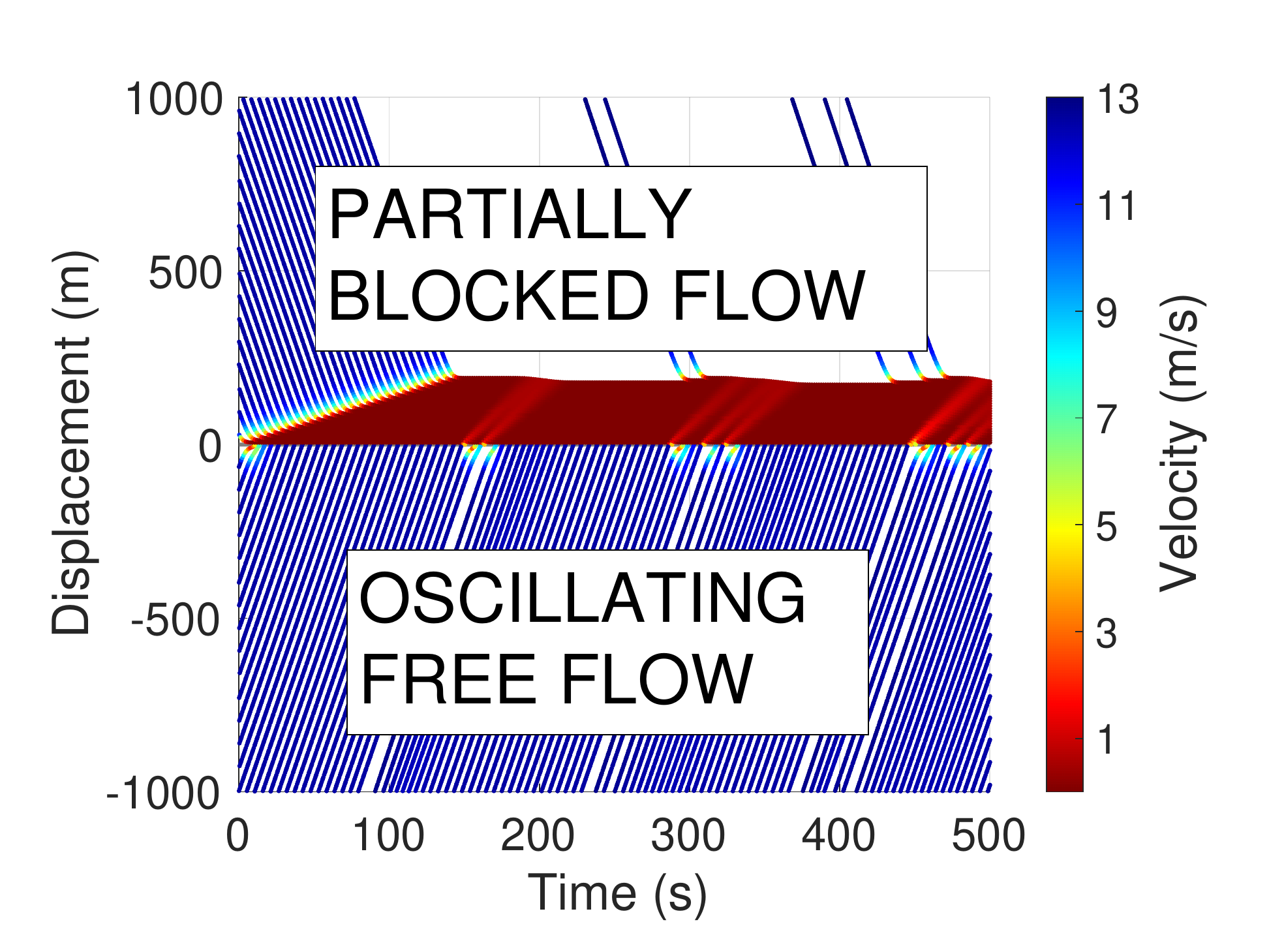}
        \subcaption{Regime (2)}
        \label{fig:e_fig}
    \end{subfigure}
        \begin{subfigure}{.32\textwidth}
        \includegraphics[width=1\textwidth]{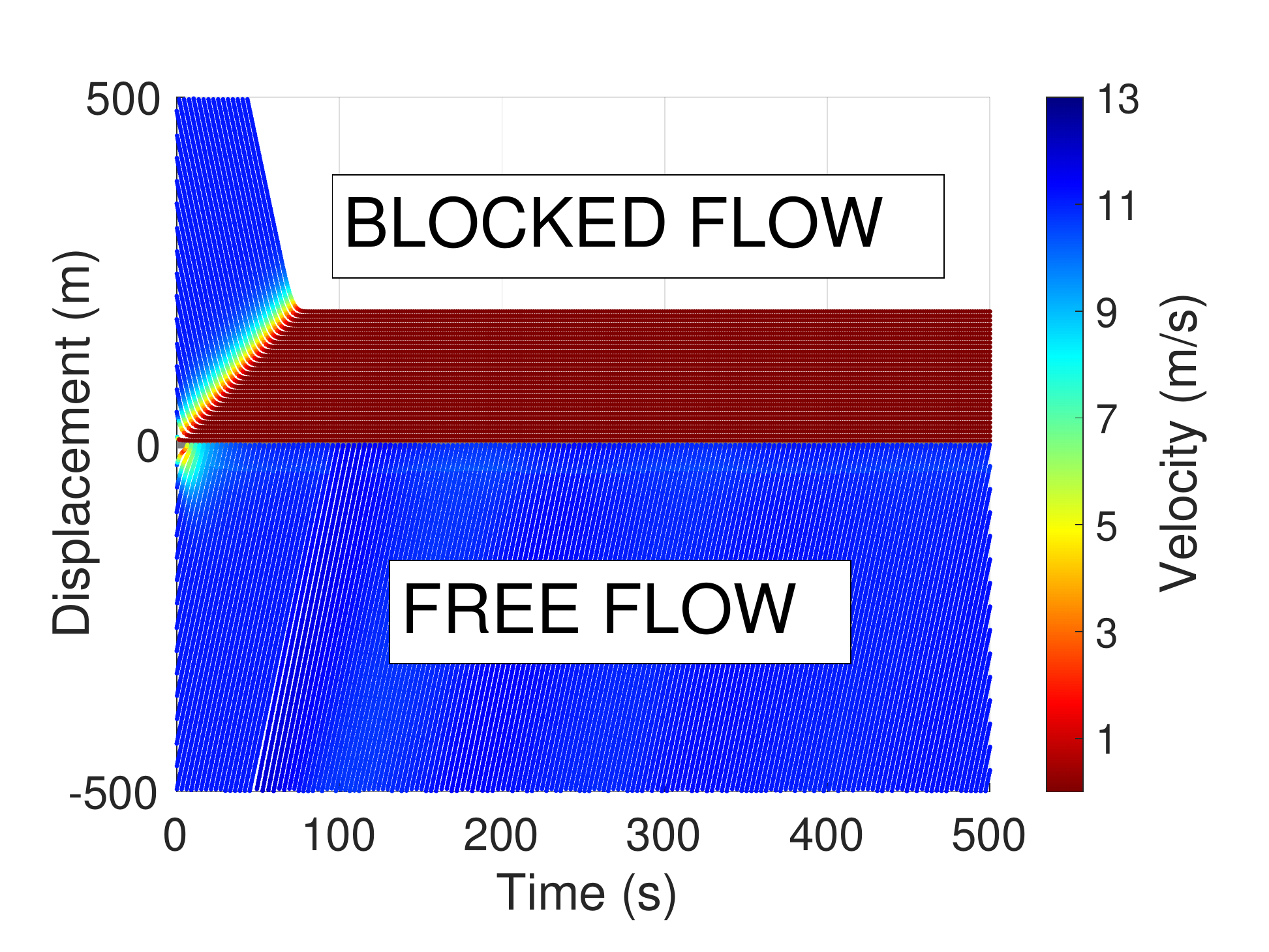}
        \subcaption{Regime (3)}
        \label{fig:c_fig}
    \end{subfigure}
    \caption{Fundamental diagram and trajectory plots for the emergent regimes}
    \label{fig:phased_a_1_5}.
\end{figure}

Results from our simulation runs are summarised in Figure~\ref{fig:a_fig}, which displays the total flow achieved across the intersection as a function of the swept density parameter. Red and green colour is used to denote crossings by the north-bound and east-bound vehicles respectively. For high densities $\rho>0.02\,\mathrm{veh/m}$ (regime (3), to the right of the dashed black line), solid blocks of colour indicate \emph{channel capture} --- that is, one arm achieves an uninterrupted flow through the intersection which entirely blocks out the other arm, whose vehicles form a stationary queue. Because of the north-east symmetry in our set-up, either arm may capture the channel in this regime, and which does so is subtly dependent on the initial data. Figure~\ref{fig:c_fig} displays vehicle trajectories upstream of the intersection for both arms in this regime. In this regime, the emergent dynamics are highly undesirable and suggest that modifications to the distributed behavioural rules are required, or potentially, the introduction of a centralised controller could be used to manage busy traffic. 

In contrast, at low densities $\rho \leq \rho_{\text{max}}$ (region (1), to the left of the dashed red line in Figure~\ref{fig:a_fig}), the alternating red/green stripes indicate that a regular turn-taking pattern emerges. In this regime, vehicles pass through the intersection without interacting with each other at all --- essentially, the junction controller is never turned on, and the flow achieves the theoretical bound of the car-following model. Figure~\ref{fig:b_fig} displays the resulting trajectories upstream of the intersection. The turn-taking pattern takes the form $\mathrm{E},\mathrm{N},\mathrm{E},\mathrm{N},\ldots$, where $\mathrm{E}$ and $\mathrm{N}$ stand for east-bound and north-bound arm vehicle crossings respectively. Note that the trajectory pattern is periodic with period $T$, where $T$ denotes the gross headway between consecutive vehicles on the same arm. Moreover, there is east-north symmetry in that the trajectory pattern is invariant to a time-shift of $T/2$ and an interchange of the east and north arms.

At intermediate densities (region (2), between the red and black dashed lines in Figure~\ref{fig:a_fig}), a variety of more complex and transient behaviours emerge. Figure~\ref{fig:e_fig} displays the trajectories. In this regime, traffic is sufficiently dense to activate the junction controller, but not dense enough to induce channel capture. Rather, one arm tends to dominate the intersection for some time, however, its stream is not regular and gaps in it allow vehicles from the other arm to cross occasionally and escape their queue. Of course, in our ring-road set-up, such vehicles rapidly rejoin the back of the queue from which they escaped. 
\section{Conclusion}
\label{sect:conclusion}
We equipped a relatively simple micro-simulation with a Behaviour Tree (BT) controller in order to study vehicle interactions at an unprioritised intersection. We discovered distinct crossing regimes in the emergent dynamics. At low densities, the junction operated efficiently and a steady turn-taking pattern emerged.  Unfortunately, at higher densities, \emph{channel capture} emerged, in which one traffic stream dominated the intersection at the expense of the other, which formed a queue. It appears to be extremely challenging to design deterministic junction controllers that are collision-free, but which guarantee good dynamics across a wide range of demand conditions.


\label{sect:bib}
\bibliographystyle{plain}
\bibliography{main}

\end{document}